\def\be{\begin{equation}}
\def\ee{\end{equation}}
\def\ba{\begin{array}}
\def\ea{\end{array}}

\documentclass[groupedaddress,aps,pra,superscriptaddress,showpacs,twocolumn,prl]{revtex4}%
\usepackage{epsfig,dsfont,amssymb,amsmath,amsthm,amsfonts,amsbsy,mathrsfs}
\usepackage{graphicx, color}
\usepackage{amsmath}
\usepackage{amssymb}
\usepackage{mathdots}
\usepackage{float}
\usepackage{cases}
\usepackage{graphpap}
\usepackage{wrapfig}
\usepackage{tikz}
\usepackage{indentfirst}
\usepackage{picinpar,graphicx}
\usepackage{graphicx}
\usepackage{dcolumn}
\usepackage{bm}
\usepackage{hyperref}

\begin{document}
\title{Near-Optimal Variance-Based Uncertainty Relations}
\author{Yunlong Xiao} 
\affiliation{Nanyang Quantum Hub, School of Physical and Mathematical Sciences, Nanyang Technological University, Singapore 637371, Singapore}
\affiliation{Max Planck Institute for Mathematics in the Sciences, 04103 Leipzig, Germany}
\author{Naihuan Jing}
\thanks{Corresponding author: jing@ncsu.edu}
\affiliation{Department of Mathematics, North Carolina State University, Raleigh, NC 27695, USA}
\affiliation{School of Mathematics, South China University of Technology, Guangzhou, Guangdong 510640, China}
\author{Bing Yu} 
\affiliation{School of Mathematics and Systems Science, Guangdong Polytechnic Normal University, Guangzhou,510665,China}
\author{Shao-Ming Fei} 
\affiliation{School of Mathematical Sciences, Capital Normal University, Beijing 100048, China}
\affiliation{Max Planck Institute for Mathematics in the Sciences, 04103 Leipzig, Germany}
\author{Xianqing Li-Jost} 
\affiliation{Max Planck Institute for Mathematics in the Sciences, 04103 Leipzig, Germany}

\begin{abstract}
Learning physical properties of a quantum system is essential for the developments of quantum technologies. However, Heisenberg's uncertainty principle constrains the potential knowledge one can simultaneously have about a system in quantum theory. Aside from its fundamental significance, the mathematical characterization of this restriction, known as `uncertainty relation', plays important roles in a wide range of applications, stimulating the formation of tighter uncertainty relations. In this work, we investigate the fundamental limitations of variance-based uncertainty relations, and introduce several `near optimal' bounds for incompatible observables. Our results consist of two morphologically distinct phases: lower bounds that illustrate the uncertainties about measurement outcomes, and the upper bound that indicates the potential knowledge we can gain. Combining them together leads to an \emph{uncertainty interval}, which captures the essence of uncertainties in quantum theory. Finally, we have detailed how to formulate lower bounds for product-form variance-based uncertainty relations by employing entropic uncertainty relations, and hence built a link between different forms of uncertainty relations.
\end{abstract}


\maketitle

\section{\uppercase\expandafter{\romannumeral1}. Introduction}
Uncertainty principle, originally introduced by Heisenberg~\cite{Heisenberg1927}, clearly sets quantum theory apart from our classical world. Formally, it states that it is impossible to predict the outcomes of incompatible measurements simultaneously, such as the position and momentum of a particle. The corresponding mathematical formulation for position and momentum are given by Kennard in Ref.~\cite{Kennard1927} (see also Ref.~\cite{Weyl1927}). Later, a general form of uncertainty relation has been established by Robertson~\cite{PhysRev.34.163}, and has been further improved by Schr\"{o}dinger in Ref.~\cite{Schrodinger1930}, which is expressed in terms of commutator and anticommutator of obserables:
\begin{align}\label{Schrodinger}
V(A)V(B)\geq|\frac{1}{2}\langle[A, B]\rangle|^{2}+|\frac{1}{2}\langle\{\overline{A},\overline{B}\}\rangle|^{2},
\end{align}
where the quantity $V(A)= \langle\overline{A}^2\rangle$ (resp. $V(B)$) stands for the variance of observable $A$ (resp. $B$), the operator $\overline{A}$ is defined as $A-\langle A\rangle$, and the expectation value $\langle~\rangle$ is over the quantum state $|\Psi\rangle$. Another way to demonstrate the joint uncertainty associated with incompatible observables is through the summation, namely $V(A)+V(B)$~\cite{PATI2007177,Xiao2016W,XiaoPhD,PhysRevA.100.032118}, which highlights an advantage in the parameter estimation of quantum system \cite{doi:10.1126/science.1104149,PhysRevLett.96.010401,PhysRevLett.100.220501,Giovannetti2011}.

Riding the waves of information theory, entropies have been used to quantify the uncertainties associated with quantum measurements~\cite{Wehner_2010}. For instance, the entropies of probability distributions of canonically conjugate variables obey Bia\l ynicki-Birula-Mycielski uncertainty relation~\cite{CMP1975}. It is noteworthy that Heisenberg's uncertainty relation follows from Ref.~\cite{CMP1975} as a special case. The entropic uncertainty relation for any pair of bounded observables is established by Deutsch in Ref.~\cite{PhysRevLett.50.631}. An improved expression was subsequently conjectured by Kraus~\cite{PhysRevD.35.3070} and then had been proved by Maassen and Uffink~\cite{PhysRevLett.60.1103}. With access to a memory system, the conventional entropic uncertainty relations have been further generalized to entanglement-assisted formalism~\cite{Berta2010}. Soon afterwards, several improvements and extensions, including the cases of multiple measurements, universal uncertainty regions and quantum processes, have been proposed in Refs.~\cite{PhysRevA.89.022112,PhysRevA.93.042125,Xiao_2016,Xiao2016U,xiao2019complementary,PhysRevResearch.3.023077}. Recently, beyond inertial frames, the uncertainty trade-off occurred near the event horizon of a Schwarzschild black hole~\cite{Huang2018} and the relativistic protocol of an uncertainty game in the presence of localized fermionic quantum fields inside cavities~\cite{PhysRevD.102.096009} have also been demonstrated.

Aside from their theoretical significance~\cite{1580791}, these uncertainty relations support a variety of applications and have been widely used in current quantum technologies, such as analyzing the security of quantum key distribution protocols~\cite{Berta2010}, witnessing quantum correlations~\cite{PhysRevA.68.032103,PhysRevLett.118.020402,Xiao_2020,RevModPhys.89.015002}, and even inferring causality from quantum dynamics~\cite{Xiao2022}. Thus, pushing the boundary of uncertainty relation will not only deepen our understanding of quantum foundations, but also has impact on practical applications.

In this work, we focus on the case of variance-based uncertainty relations, with the forms of both product and summation, and introduce the concept of uncertainty interval. The formulation of such an interval can of course be subdivided into two, namely finding the lower bound and upper bounds for joint uncertainties. To do so, we establish the \emph{partial Cauchy-Schwarz inequality}, which generalizes the standard Cauchy-Schwarz inequality, and use this toolkit to construct near-optimal bounds for variance-based uncertainty relations. Numerical results highlight the advantages of our framework.

\section{\uppercase\expandafter{\romannumeral2}. Product-Form Variance-Based Uncertainty Relations}

Throughout this paper, we consider quantum systems acting on finite-dimensional Hilbert space. Let us start with a pair of incompatible observables $A$ and $B$, and denote their spectral decompositions as $A=\sum_{i}a_{i}|a_{i}\rangle\langle a_{i}|$ and $B=\sum_{i}b_{i}|b_{i}\rangle\langle b_{i}|$ respectively. On the other hand, assume the alternative observable $\overline{A}$ and $\overline{B}$ have the following spectral decompositions; that are $\overline{A}=\sum_{i}{a}_i'|a_{i}\rangle\langle a_{i}|$ and $\overline{B}=\sum_{i}{b}_i'|b_{i}\rangle\langle b_{i}|$. Remark that, here all the eigenvalues are real numbers, i.e. $a_{i}, {a}_i', b_{i}, {b}_i'\in\mathbb{R}$. Now for any given orthonormal basis $\{|\psi_{i}\rangle\}$, we can re-express $\overline{A}|\Psi\rangle$ and $\overline{B}|\Psi\rangle$ as $\sum_{i}\alpha_{i}|\psi_{i}\rangle$ and $\sum_{i}\beta_{i}|\psi_{i}\rangle$ respectively. It is worth mentioning that in general both $\overline{A}|\Psi\rangle$ and $\overline{B}|\Psi\rangle$ are unnormalized, and hence the vectors $(\alpha_{i})$ and $(\beta_{i})$ do not forms probability distributions. Then, by defining the absolute value of $\alpha_{i}$ and $\beta_{i}$ as $x_i$ and $y_i$ respectively, the variance of observables $A$ and $B$ can be rewritten as
\begin{align}
V(A)=|\overrightarrow{x}|^{2},\quad
V(B)=|\overrightarrow{y}|^{2},
\end{align}
and thus we have
\begin{align}
V(A)V(B)= |\overrightarrow{x}|^{2}\cdot|\overrightarrow{y}|^{2}.
\end{align}
It now follows from Cauchy-Schwarz inequality immediately that
\begin{align}\label{e:cs}
 V(A)V(B)\geq\left(\sum\limits_{i}x_{i}y_{i}\right)^{2}.
\end{align}
We note that such a choice of $x_i$ and $y_i$ leads directly to the main results presented in a recent formulation of strong uncertainty relation~\cite{PhysRevA.95.052117}. Clearly, this is not the only choice of $x_i$ and $y_i$. By setting $x_{i}$ as $|{a}_i'|\sqrt{\langle\Psi|a_{i}\rangle\langle a_{i}|\Psi\rangle}$ and $y_{i}$ as $|{b}_i'|\sqrt{\langle\Psi|b_{i}\rangle\langle b_{i}|\Psi\rangle}$, we re-obtain another part of results constructed in Ref.~\cite{PhysRevA.95.052117}. Here, for simplicity, we further denote the Uhlmann's fidelity between $|\Psi\rangle$ and $|a_{i}\rangle$ ($|b_{i}\rangle$) as $F^{a}_{i}$ ($F^{b}_{i}$), which are
\begin{align}
F^{a}_{i}
=
\langle\Psi|a_{i}\rangle\langle a_{i}|\Psi\rangle,\quad
F^{b}_{i}
=
\langle\Psi|b_{i}\rangle\langle b_{i}|\Psi\rangle.
\end{align}

A key observation in this work is that any improvement over the well-known Cauchy-Schwarz inequality will give us a better bound of variance-based uncertainty relation, with the same amount of information required in Eq.~\ref{e:cs}. To this end, we investigate the intrinsic connection between the arithmetic-geometric mean (AM-GM) inequality and the Cauchy-Schwarz inequality. We start by writing down the product of $|\overrightarrow{\alpha}|^{2}$ and $|\overrightarrow{\beta}|^{2}$,
\begin{align}
|\overrightarrow{\alpha}|^{2}|\overrightarrow{\beta}|^{2}=\sum\limits_{ij}x_{i}^{2}y_{j}^{2}
 &=\sum\limits_{i<j}(x_{i}^{2}y_{j}^{2}+x_{j}^{2}y_{i}^{2})+\sum\limits_{i}x_{i}^{2}y_{i}^{2}\notag\\
 &\geq\sum\limits_{i<j}(2x_{i}x_{j}y_{j}y_{i})+\sum\limits_{i}x_{i}^{2}y_{i}^{2}\notag\\
 &=\left(\sum\limits_{i}x_{i}y_{i}\right)^{2}.
\end{align}
Above inequality is a result of $n(n-1)/2$ rounds of AM-GM inequalities for $x_{i}^2y_{j}^2+x_{j}^2y_{i}^2\geq 2x_iy_jx_jy_i$ with different indexes. Therefore,
the equality condition holds if and only if $x_{i}y_{j}=x_{j}y_{i}$ for all $i\neq j$. By defining the quantity $I_{k}$ as
\begin{align}\label{e:ik}
 \sum\limits_{1\leq i<j\leq k}(2x_{i}x_{j}y_{j}y_{i})+&\sum\limits_{\substack{1\leq i<j\leq n \\ k<j}}(x_{i}^{2}y_{j}^{2}+x_{j}^{2}y_{i}^{2})+\sum\limits_{1\leq i\leq n}x_{i}^{2}y_{i}^{2},
\end{align}
we can write the left-hand-side of Eq.~\ref{e:cs} as
\begin{align}
I_{0}=|\overrightarrow{x}|^{2}|\overrightarrow{y}|^{2}=V(A)V(B),
\end{align}
which is precisely the product-form joint uncertainty. On the other hand, the previous known bound in Ref.~\cite{PhysRevA.95.052117}, i.e. right-hand-side quantity of Eq.~\ref{e:cs}, can be reformatted as
\begin{align}
I_{n}=\left(\sum\limits_{i}x_{i}y_{i}\right)^{2}.
\end{align}
Now we introduce a chain of inequalities that outperform Cauchy-Schwarz inequality. More precisely, we have

\noindent\textbf{Theorem 1.} {\it For any $n$-dimensional real vectors $\overrightarrow{x}$,
$\overrightarrow{y}$ with non-negative components, and $I_k$ defined in Eq.~\ref{e:ik}, we have}
\begin{align}\label{e:pcs}
I_{0}\geq I_{2}\geq\cdots\geq I_{n-1}\geq I_{n}.
\end{align}
Actually, for any index $k$ it follow from the AM-GM inequality that
\begin{align}
I_{k+1}=I_k+\sum_{i=1}^{k}(2x_ix_{k+1}y_iy_{k+1}-x_i^2y_{k+1}^2-x_{k+1}^2y_i^2)\leq I_k,
\end{align}
as required. Algebraically, the inequality $|\overrightarrow{x}|^{2}|\overrightarrow{y}|^{2}\geq I_{k}$ is obtained by applying AM-GM inequality to the first $k$ components of both $\overrightarrow{x}$ and
$\overrightarrow{y}$, and hence can be viewed as a partial Cauchy-Schwarz inequality. More importantly, such a partial Cauchy-Schwarz inequality, see Eq.~\ref{e:pcs}, provides
$n-2$ tighter lower bounds for $V(A)V(B)$ compared with the main result of \cite{PhysRevA.95.052117}, namely $I_0=V(A)V(B)\geq I_n$. In particular, we can insert more terms in the above descending chain by selecting arbitrary $x_{i}^{2}y_{j}^{2}+x_{j}^{2}y_{i}^{2}$ ($i<j$). For example,
the inequality $I_0\geq I_{n-1}$ obtained from our Thm.~1 immediately leads to a tighter bound. More precisely, Eq.~\ref{e:cs} can be improved to
\begin{align}\label{e:l1}
 V(A)V(B)\geq&\frac{1}{4}\left(\sum\limits_{i=1}^{n-1}\left|\langle[\overline{A}, \overline{B}_{n}]\rangle+\langle\{\overline{A}, \overline{B}_{n}\}\rangle\right|\right)^{2}\notag\\
 +&\left|\langle\Psi|\overline{A}|\psi_{n}\rangle\right|^{2}(\sum\limits_{i=1}^{n}\left|\langle\Psi|\overline{B}|\psi_{n}\rangle\right|^{2})\notag\\
 +&\left|\langle\Psi|\overline{B}|\psi_{n}\rangle\right|^{2}(\sum\limits_{i=1}^{n}\left|\langle\Psi|\overline{A}|\psi_{n}\rangle\right|^{2})\notag\\
 -&\left|\langle\Psi|\overline{A}|\psi_{n}\rangle\right|^{2}\left|\langle\Psi|\overline{B}|\psi_{n}\rangle\right|^{2}:=\mathcal{L}_{1},
\end{align}
which offers a stronger bound than that of
\begin{align}
 \mathcal{L}_{1}\geq\frac{1}{4}\left(\sum\limits_{i=1}^{n}\left|\langle[\overline{A}, \overline{B}_{n}]\rangle+\langle\{\overline{A}, \overline{B}_{n}\}\rangle\right|\right)^{2}\geq
 \left|\langle\overline{A}\overline{B}\rangle\right|^{2}.
\end{align}
Note that the method of constructing bounds presented here for variance-based uncertainty relations requires the same amount of information, i.e. the fidelity between quantum state and the eigenvector of observables, needed in previous works, such as the one considered in Ref.~\cite{PhysRevA.95.052117}, but provable tighter.

We now move on to further strengthening the bounds of uncertainty relations by considering the action of symmetric group $\mathfrak{S}_{n}$. For any two permutations $\pi_{1}, \pi_{2}\in\mathfrak{S}_{n}$, we define
\begin{align}
 (\pi_{1}, \pi_{2})I_{k}=&\sum\limits_{1\leq \pi_{1}(i)<\pi_{2}(j)\leq k}(2x_{\pi_{1}(i)}x_{\pi_{2}(j)}y_{\pi_{2}(j)}y_{\pi_{1}(i)})\notag\\
 +&\sum\limits_{\substack{1\leq \pi_{1}(i)<\pi_{2}(j)\leq n\\k<\pi_{2}(j)}}(x_{\pi_{1}(i)}^{2}y_{\pi_{2}(j)}^{2}+x_{\pi_{2}(j)}^{2}y_{\pi_{1}(i)}^{2})\notag\\
 +&\sum\limits_{\pi_{1}(i)=\pi_{2}(j)}x_{\pi_{1}(i)}^{2}y_{\pi_{2}(j)}^{2}.
\end{align}
It is straightforward to check that the quantity $I_{0}$ is stable under the action of $\mathfrak S_n\times \mathfrak S_n$. Writing everything out explicitly, we have

\noindent\textbf{Theorem 2.}  {\it For any permutations $\pi_{1}, \pi_{2}\in\mathfrak{S}_{n}$, we have}
\begin{align}
I_{0}\geq (\pi_{1}, \pi_{2})I_{2}\geq\cdots\geq (\pi_{1}, \pi_{2})I_{n-1}\geq (\pi_{1}, \pi_{2})I_{n}.
\end{align}
Optimizing over the symmetric group $\mathfrak S_n$, a stronger version of the variance-based uncertainty relations is obtained.

\noindent\textbf{Theorem 3.} {\it For any permutations $\pi_{1}, \pi_{2}\in\mathfrak{S}_{n}$, we have}
\begin{align}
I_{0}&\geq \max\limits_{\pi_{1}, \pi_{2}\in\mathfrak{S}_{n}}(\pi_{1}, \pi_{2})I_{2}\geq\cdots\geq \max\limits_{\pi_{1}, \pi_{2}\in\mathfrak{S}_{n}}(\pi_{1}, \pi_{2})I_{n}.
\end{align}
Mathematically, above inequalities are tighter than the result in Thm.~1, since $\max\limits_{\pi_{1}, \pi_{2}\in\mathfrak{S}_{n}}(\pi_{1}, \pi_{2})I_{k}\geq I_{k}$ holds for any permutations. Physically, the action of symmetric group works well since the overlaps between quantum state and the eigenvectors of observables are not uniformly distributed.

\section{\uppercase\expandafter{\romannumeral3}. Sum-Form Variance-Based Uncertainty Relations}

In this section we turn our attention to the sum-form variance-based uncertainty relations. Before doing so, let us
recall the {\it rearrangement inequality} first. Let $(x_{i})$ and ($y_{i}$) be two $n$-tuple of real positive numbers arranged in non-increasing order, namely $x_{i}\geq x_{i+1}$ and $y_{i}\geq y_{i+1}$, with their {\it direct sum}, {\it random sum} and {\it reverse sum} between $x_{i}$ and $y_{i}$ being defined as
\begin{align}
 Di:=&x_{1}y_{1}+x_{2}y_{2}+\cdots+x_{n}y_{n},\notag\\
 Ra:=&x_{1}y_{\pi(1)}+x_{2}y_{\pi(2)}+\cdots+x_{n}y_{\pi(n)},~~\pi\in\mathfrak{S}_{n}\notag\\
 Re:=&x_{1}y_{n}+x_{2}y_{n-1}+\cdots+x_{n}y_{1}.\notag\\
\end{align}
Then the following lemma characterizes the relationship among these quantities; that is

\noindent\textbf{Lemma.}({\bf Rearrangement inequality}) {\it For any two non-increasing $n$-tuples $x$ and $y$ of nonnegative numbers, we have}
\begin{align}
 Di\geq Ra\geq Re.
\end{align}
From the parallelogram law, the summation of variances can be re-expressed as
\begin{align}
V(A)+V(B)=\frac{1}{2}\sum\limits_{i}(x_{i}+y_{i})^{2}+\frac{1}{2}\sum\limits_{i}(x_{i}-y_{i})^{2}.
\end{align}
Combining with the rearrangement inequality we obtain the following result.

\noindent\textbf{Theorem 4.} {\it
For any two permutations $\pi_1, \pi_2\in\mathfrak{S}_{n}$, we have}
\begin{align}\label{e:l2}
V(A)+V(B)\geq&\frac{1}{2}\sum\limits_{i}(x_{i}+y_{i})(x_{\pi_{1}(i)}+y_{\pi_{1}(i)})\notag\\
+&\frac{1}{2}\sum\limits_{i}\left|x_{i}-y_{i}\right|\left|x_{\pi_{2}(i)}-y_{\pi_{2}(i)}\right|.
\end{align}
Remark that, by setting $\pi_{1}=(1)$, our newly constructed uncertainty relation outperforms similar results
of sum-form variance-based uncertainty relation considered in Ref.~\cite{PhysRevA.95.052117}. We denote by $\mathcal{L}_{2}$ the
bound of Thm. 4 corresponding to the choice
of
$\pi_{1}=(1)$, $\pi_{2}=(1~~2~\cdots~n)$, $x_{i}=\left|\alpha_{i}\right|$,
$y_{i}=\left|\beta_{i}\right|$, which will be used in Sec. \uppercase\expandafter{\romannumeral5}.

\section{\uppercase\expandafter{\romannumeral4}. Uncertainty Intervals}

Quantum theory does not only impose restrictions on the lower bounds of uncertainties, but also sets limitations on the upper bounds of uncertainties~\cite{PhysRevA.95.052117}, which are known as reverse uncertainty relations in the literature. In this section,
we investigate the reverse uncertainty relations for both the product-form and sum-form uncertainty relations, and introduce several tighter bounds. Consequently, our lower bounds presented in previous sections together with the results obtained in this section lead to intervals for joint uncertainty, which are referred as uncertainty intervals.

For index $1\leq i\leq n$, we define
\begin{align}
X&=\max\limits_{i}\{x_{i}\},\quad x=\min\limits_{i}\{x_{i}\},\notag\\
Y&=\max\limits_{i}\{y_{i}\},\quad\, y=\min\limits_{i}\{y_{i}\}.
\end{align}
Using the rearrangement inequality, we thus see that
\begin{align}\label{e:up11}
 \frac{(xy+XY)^{2}}{4xyXY}\left(\sum\limits_{i}x_{i}y_{i}\right)^{2}&\geq\frac{(xy+XY)^{2}}{4xyXY}\left(\sum\limits_{i}x_{i}y_{\pi(i)}\right)^{2}\notag\\
 &\geq V(A)V(B).
\end{align}
By taking minimum over all permutations $\pi\in\mathfrak{S}_{n}$, we obtain a tighter upper bound for $V(A)V(B)$:
\begin{align}\label{e:up12}
 V(A)V(B)\leq\min\limits_{\pi\in\mathcal{S}_{n}}\frac{(xy+XY)^{2}}{4xyXY}\left(\sum\limits_{i}x_{i}y_{\pi(i)}\right)^{2}:=\mathcal{U}_{1},
\end{align}
which proves that the joint uncertainty of incompatible observables $A$ and $B$ (for the
product-form) is restricted within the interval
$[\mathcal{L}_{1}, \mathcal{U}_{1}]$, i.e. $V(A)V(B)\in[\mathcal{L}_{1}, \mathcal{U}_{1}]$. In other words, $[\mathcal{L}_{1}, \mathcal{U}_{1}]$ is an {\it uncertainty interval} for $V(A)V(B)$.


\begin{figure}
\centering
\includegraphics[width=0.48\textwidth]{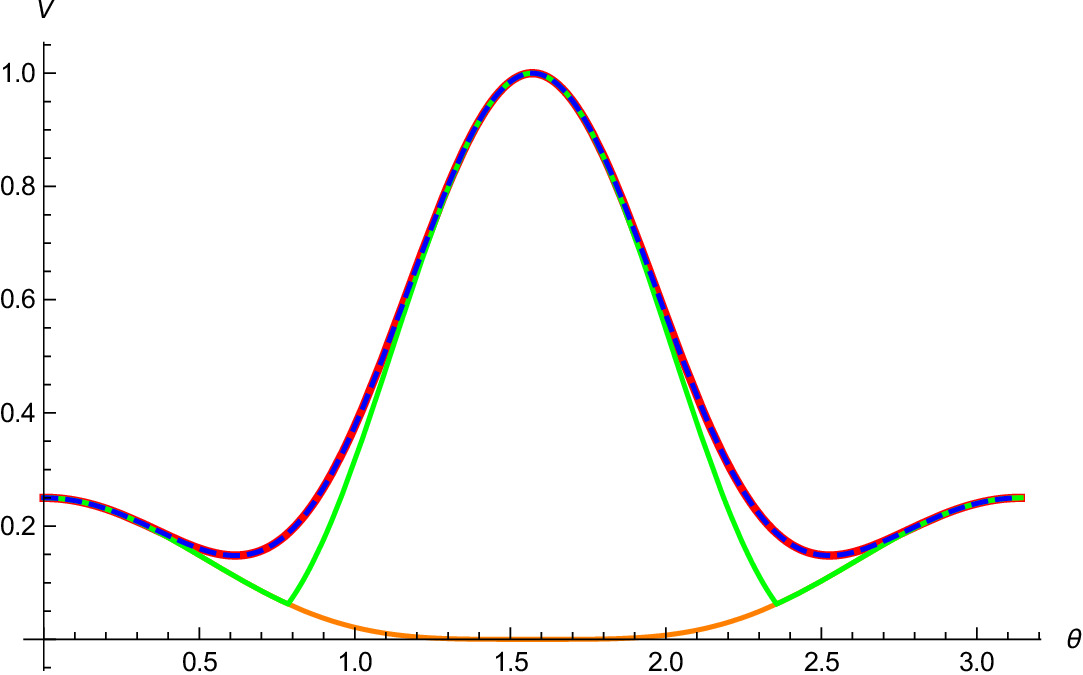}
\caption{Lower bounds of $V(A)V(B)$ for a family of spin-$1$ particles $|\Psi(\theta)\rangle$: the product-form uncertainty relation $V(A)V(B)$, the bound $\mathcal{L}_{1}$ of Eq.~\ref{e:l1}, the bound of Ref.~\cite{PhysRevA.95.052117}, and the bound of Schr\"{o}dinger uncertainty relation~\cite{Schrodinger1930} are depicted in red, blue, green, and orange respectively.
}
\label{Figure1}
\end{figure}

On the other hand, using the fact $V(A)=|\overrightarrow{\alpha}|^{2}$ and $V(B)=|\overrightarrow{\beta}|^{2}$,
one derive an upper bound on the sum of variances of incompatible observables $A$ and $B$ as
\begin{align}\label{e:up21}
 V(A)+V(B)=\sum\limits_{i}(x_{i}^{2}+y_{i}^{2})\leq\sum\limits_{i}(x_{i}+y_{i})^{2}.
\end{align}
Recalling the definitions $x_{i}=\left|\alpha_{i}\right|$ and $y_{i}=\left|\beta_{i}\right|$, we have that
\begin{align}\label{e:up22}
 V(A)+V(B)\leq\sum\limits_{i}(\left|\langle\psi_{n}|\overline{A}|\Psi\rangle\right|+\left|\langle\psi_{n}|\overline{B}|\Psi\rangle\right|)^{2}.
\end{align}
Denote the right-hand (RHS) of (\ref{e:up22}) by $\mathcal{U}_{2}$. Thus we have obtained
a uncertainty interval for $V(A)+V(B)$: $[\mathcal{L}_{2}, \mathcal{U}_{2}]$. We remark that $\mathcal{U}_{2}$ is not always better than the bound obtained by
~\cite{PhysRevA.95.052117}, but it provides a complementary one. The comparison will be discussed
by examples in the next section.

\begin{figure}
\centering
\includegraphics[width=0.48\textwidth]{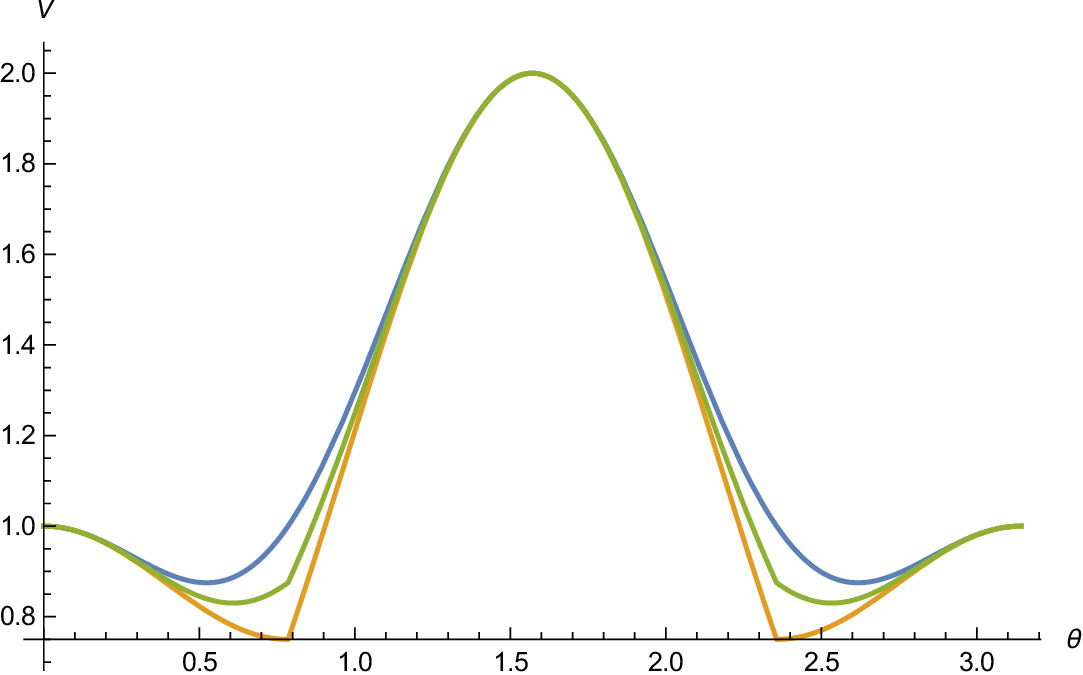}
\caption{Lower bounds of $V(A)+V(B)$ for a family of spin-$1$ particles $|\Psi(\theta)\rangle$: the sum-form uncertainty relation $V(A)+V(B)$, our bound $\mathcal{L}_{2}$ of Eq.~\ref{e:l2}, and the bound of Ref.~\cite{PhysRevA.95.052117} are depicted in blue, green, and yellow respectively.}
\label{Figure2}
\end{figure}

\section{\uppercase\expandafter{\romannumeral5}. Numerical Examples and Conclusion}
In this section we provide numerical examples to show how the bounds obtained in this work outperform previous strong results~\cite{PhysRevA.95.052117}. First of all, let us consider the spin-$1$ particle with the state $|\Psi(\theta)\rangle=\cos\theta|1\rangle-\sin\theta|0\rangle$, where the state $|0\rangle$ and $|1\rangle$ are eigenstates of the angular momentum $L_{z}$. We investigate the uncertainty associated with angular momentum operators for spin-$1$ particle, namely $A=L_{x}$ and $B=L_{y}$. To formulate bounds for uncertainty relations, we choose $x_{i}=\left|\alpha_{i}\right|$ and $y_{i}=\left|\beta_{i}\right|$ (similar for
$x_{i}=|{a}_i'|\sqrt{\langle\Psi(\theta)|a_{i}\rangle\langle a_{i}|\Psi(\theta)\rangle}$ and $y_{i}=|{b}_i'|\sqrt{\langle\Psi(\theta)|b_{i}\rangle\langle b_{i}|\Psi(\theta)\rangle}$).

In Fig.~\ref{Figure1}, our bound $\mathcal{L}_{1}$ has been compared with that of~\cite{PhysRevA.95.052117} in the product-form for the family of spin-$1$ particles $|\Psi(\theta)\rangle$. As shown in our numerical results, the bound $\mathcal{L}_{1}$ (in blue) provides the best estimation and is almost optimal. As a supplement, we also compare our result with Schr\"odinger's uncertainty relation (in orange). In Fig.~\ref{Figure2}, we plot lower bounds for the sum-form variance-based uncertainty relation for the family of the spin-$1$ particles $|\Psi(\theta)\rangle$, which highlights the advantage of our method.

\begin{figure}
\centering
\includegraphics[width=0.48\textwidth]{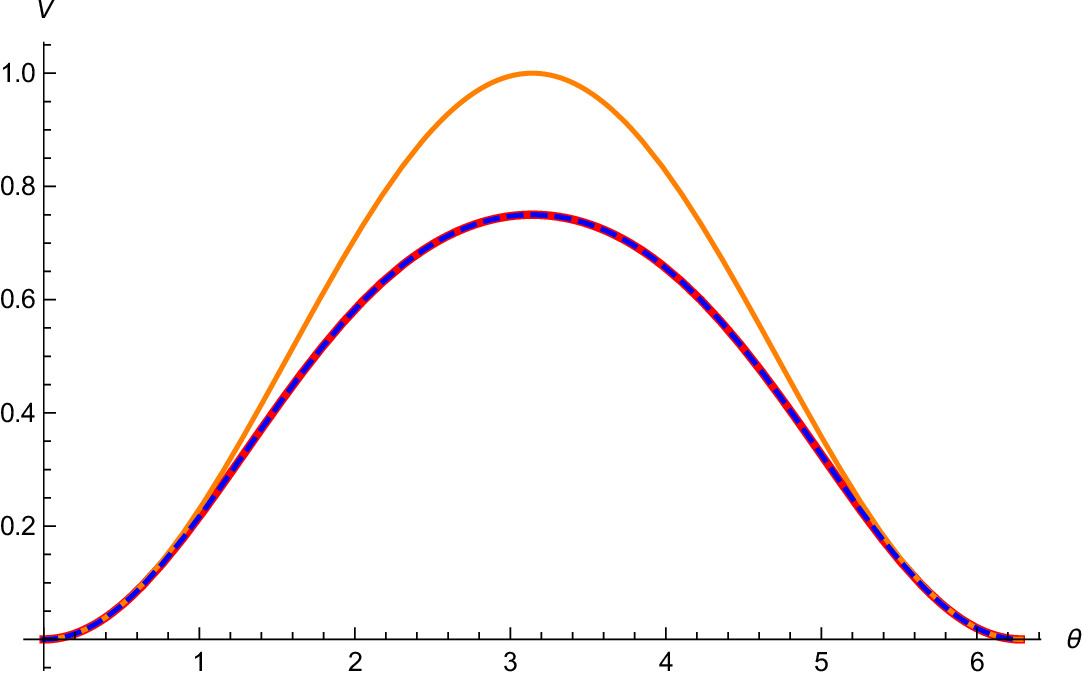}
\caption{Upper bounds of $V(A)V(B)$ for a family of spin-$1/2$ particles $\rho(\theta)$:
the product-form uncertainty relation $V(A)V(B)$, our bound $\mathcal{U}_{1}$ of Eq.~\ref{e:up12}, and the bound of Ref.~\cite{PhysRevA.95.052117} are depicted in red, blue, and orange respectively.}
\label{Figure3}
\end{figure}

\begin{figure}
\centering
\includegraphics[width=0.48\textwidth]{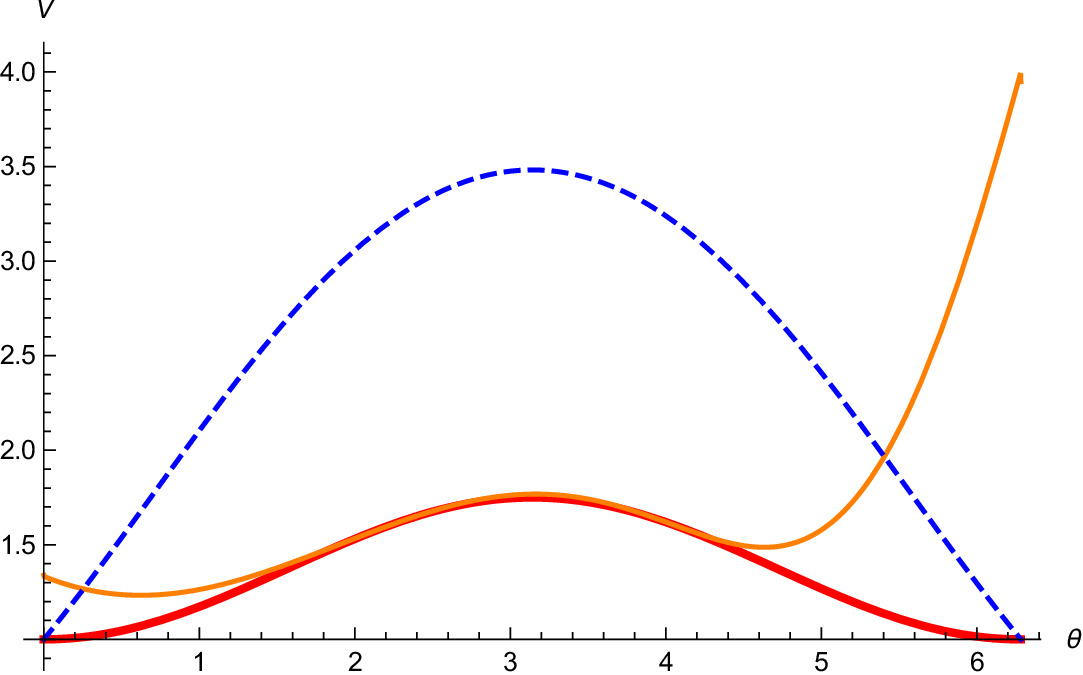}
\caption{Upper bounds of $V(A)+ V(B)$ for a family of spin-$1/2$ particles $\rho(\theta)$:
the sum-form uncertainty relation $V(A)V(B)$, our bound $\mathcal{U}_{2}$ of Eq.~\ref{e:up22}, and the bound of Ref.~\cite{PhysRevA.95.052117} are depicted in red, blue, and orange respectively.}
\label{Figure4}
\end{figure}

Let us move on to considering the spin-$\frac{1}{2}$ particle with the following density matrix
\begin{align}
\rho(\theta)=\frac{1}{2}\left(Id+\cos\frac{\theta}{2}\sigma_{x}+\frac{\sqrt{3}}{2}\sin\frac{\theta}{2}\sigma_{y}+\frac{1}{2}\sin\frac{\theta}{2}\sigma_{z}\right),
\end{align}
where the two incompatible observables are taken as $A=\sigma_{x}$ and $B=\sigma_{z}$. In Fig.~\ref{Figure3},
it has been shown that our upper bound $\mathcal{U}_{1}$ provides the best estimation for the product of two variances and typically outperforms the upper bound from Ref.~\cite{PhysRevA.95.052117}. Note that our bound is almost optimal, as it is almost identical to the optimal value. However, our upper bound $\mathcal{U}_{2}$ for the sum of variances $V(A)+V(B)$ for states $\rho(\theta)$ is not always tighter than that of Ref.~\cite{PhysRevA.95.052117}. Nevertheless, it still provides an improvements for most of the time. See Fig.~\ref{Figure4} for an illustration.

Apart from constructing stronger uncertainty relations, our method introduced in Sec. \uppercase\expandafter{\romannumeral2} also helps to fill up the gap between product-form variance-based uncertainty relations and entropic uncertainty relations. Following Ref.~\cite{PhysRevA.86.024101}, we have
\begin{align}
 V(A)+V(B)\geq H(A)+H(B)+c,
\end{align}
where $H(\cdot)$ stands for the Shannon entropy and $c$ is a state-independent constant. Using Thm.~1, it is straightforward to check that
\begin{align}\label{e:entropy1}
 V(A)V(B)
 \geq\frac{1}{4}\left(\sum\limits_{i=1}^{n-1}x_{i}y_{i}\right)^{2}
 +x_{n}^{2}V(B)+y_{n}^{2}V(A)-x_{n}^{2}y_{n}^{2}.
\end{align}
On the one hand, the term $x_{n}^{2}V(B)+y_{n}^{2}V(A)$ appeared above forms a so-called
{\it weighted uncertainty relation}~\cite{Xiao2016W}. Notice that we can always assume $x_{n}^{2}=y_{n}^{2}$ in the numerical calculation, since $V(rA)V(B)=r^{2}V(A)V(B)$. Thus, Eq.~\ref{e:entropy1} can be bounded as
\begin{align}\label{e:entropy2}
 V(A)V(B)
 \geq\frac{1}{4}\left(\sum\limits_{i=1}^{n-1}x_{i}y_{i}\right)^{2}
 +x_{n}^{2}\left(H(A)+H(B)+c\right)-x_{n}^{4}.
\end{align}
Therefore both the incompatibility between observables and mixness of the quantum state will affect
the variance-based uncertainty relations. Moreover, any entropic uncertainty relation can be employed to construct a lower bound for product-form variance-based uncertainty relation.

To summarize, we have introduced several variance-based uncertainty relations both in the sum and product forms. Our results contain both the lower bounds and the upper bounds, which leads to the concept of uncertainty intervals. Numerical experiments illustrate the advantages of our bounds, and in some cases our bounds are near optimal. Quite remarkable, our method in deriving stronger variance-based uncertainty relations also fills the gap between the product-form variance-based uncertainty relations and the entropic uncertainty relations. Beside the results present here, our framework can also be used in formulating unitary uncertainty relations. For more details, see our follow-up work~\cite{PhysRevA.100.022116}.

\smallskip
\noindent{\bf Acknowledgments}\, \,  Y. Xiao is supported by the Natural Sciences, the National Research Foundation (NRF). Singapore, under its NRFF Fellow programme (Grant No. NRF-NRFF2016-02), Singapore Ministry of Education Tier 1 Grants RG162/19 (S), the Quantum Engineering Program QEP-SF3, and No FQXi-RFP-1809 (The Role of Quantum Effects in Simplifying Quantum Agents) from the Foundational Questions Institute and Fetzer Franklin Fund (a donor-advised fund of Silicon Valley Community Foundation). B. Yu acknowledges the support of Startup Funding of Guangdong Polytechnic Normal University No. 2021SDKYA178, and Guangdong Basic and Applied Basic Research Foundation No. 2020A1515111007. S.-M. Fei acknowledges the support of National Natural Science Foundation of China (NSFC) under Grant Nos. 12075159 and 12171044; Beijing Natural Science Foundation (Grant No. Z190005); the Academician Innovation Platform of Hainan Province. The work is supported by National Natural Science Foundation of China (grant Nos. 12126351 and 12126314), Natural Science Foundation of Hubei Province grant No. 2020CFB538, China Scholarship Council and Simons Foundation grant No. 523868. Any opinions, findings and conclusions or recommendations expressed in this material are those of the author(s) and do not reflect the views of National Research Foundation or Monistry of Education, Singapore.
\bibliography{Bib}

\begin{thebibliography}{36}
\expandafter\ifx\csname natexlab\endcsname\relax\def\natexlab#1{#1}\fi
\expandafter\ifx\csname bibnamefont\endcsname\relax
  \def\bibnamefont#1{#1}\fi
\expandafter\ifx\csname bibfnamefont\endcsname\relax
  \def\bibfnamefont#1{#1}\fi
\expandafter\ifx\csname citenamefont\endcsname\relax
  \def\citenamefont#1{#1}\fi
\expandafter\ifx\csname url\endcsname\relax
  \def\url#1{\texttt{#1}}\fi
\expandafter\ifx\csname urlprefix\endcsname\relax\def\urlprefix{URL }\fi
\providecommand{\bibinfo}[2]{#2}
\providecommand{\eprint}[2][]{\url{#2}}

\bibitem[{\citenamefont{Heisenberg}(1927)}]{Heisenberg1927}
\bibinfo{author}{\bibfnamefont{W.}~\bibnamefont{Heisenberg}},
  \bibinfo{journal}{Zeitschrift f{\"u}r Physik} \textbf{\bibinfo{volume}{43}},
  \bibinfo{pages}{172} (\bibinfo{year}{1927}),
  \urlprefix\url{https://doi.org/10.1007/BF01397280}.

\bibitem[{\citenamefont{Kennard}(1927)}]{Kennard1927}
\bibinfo{author}{\bibfnamefont{E.~H.} \bibnamefont{Kennard}},
  \bibinfo{journal}{Zeitschrift f{\"u}r Physik} \textbf{\bibinfo{volume}{44}},
  \bibinfo{pages}{326} (\bibinfo{year}{1927}),
  \urlprefix\url{https://doi.org/10.1007/BF01391200}.

\bibitem[{\citenamefont{Weyl}(1927)}]{Weyl1927}
\bibinfo{author}{\bibfnamefont{H.}~\bibnamefont{Weyl}},
  \bibinfo{journal}{Zeitschrift f{\"u}r Physik} \textbf{\bibinfo{volume}{46}},
  \bibinfo{pages}{1} (\bibinfo{year}{1927}),
  \urlprefix\url{https://doi.org/10.1007/BF02055756}.

\bibitem[{\citenamefont{Robertson}(1929)}]{PhysRev.34.163}
\bibinfo{author}{\bibfnamefont{H.~P.} \bibnamefont{Robertson}},
  \bibinfo{journal}{Phys. Rev.} \textbf{\bibinfo{volume}{34}},
  \bibinfo{pages}{163} (\bibinfo{year}{1929}),
  \urlprefix\url{https://link.aps.org/doi/10.1103/PhysRev.34.163}.

\bibitem[{\citenamefont{Schr\"odinger}(1930)}]{Schrodinger1930}
\bibinfo{author}{\bibfnamefont{E.}~\bibnamefont{Schr\"odinger}},
  \bibinfo{journal}{Ber. Kgl. Akad. Wiss. Berlin}
  \textbf{\bibinfo{volume}{24}}, \bibinfo{pages}{296} (\bibinfo{year}{1930}).

\bibitem[{\citenamefont{Pati and Sahu}(2007)}]{PATI2007177}
\bibinfo{author}{\bibfnamefont{A.}~\bibnamefont{Pati}} \bibnamefont{and}
  \bibinfo{author}{\bibfnamefont{P.}~\bibnamefont{Sahu}},
  \bibinfo{journal}{Physics Letters A} \textbf{\bibinfo{volume}{367}},
  \bibinfo{pages}{177} (\bibinfo{year}{2007}), ISSN \bibinfo{issn}{0375-9601},
  \urlprefix\url{https://www.sciencedirect.com/science/article/pii/S0375960107003696}.

\bibitem[{\citenamefont{Xiao et~al.}(2016{\natexlab{a}})\citenamefont{Xiao,
  Jing, Li-Jost, and Fei}}]{Xiao2016W}
\bibinfo{author}{\bibfnamefont{Y.}~\bibnamefont{Xiao}},
  \bibinfo{author}{\bibfnamefont{N.}~\bibnamefont{Jing}},
  \bibinfo{author}{\bibfnamefont{X.}~\bibnamefont{Li-Jost}}, \bibnamefont{and}
  \bibinfo{author}{\bibfnamefont{S.-M.} \bibnamefont{Fei}},
  \bibinfo{journal}{Scientific Reports} \textbf{\bibinfo{volume}{6}},
  \bibinfo{pages}{23201} (\bibinfo{year}{2016}{\natexlab{a}}),
  \urlprefix\url{https://doi.org/10.1038/srep23201}.

\bibitem[{\citenamefont{Xiao}(2017)}]{XiaoPhD}
\bibinfo{author}{\bibfnamefont{Y.}~\bibnamefont{Xiao}}, Ph.D. thesis,
  \bibinfo{school}{Leipzig University}, \bibinfo{address}{Leipzig, Germany}
  (\bibinfo{year}{2017}),
  \urlprefix\url{https://ul.qucosa.de/landing-page/?tx_dlf[id]=https%3A%2F%2Ful.qucosa.de%2Fapi%2Fqucosa%253A15366%2Fmets}.

\bibitem[{\citenamefont{Xiao et~al.}(2019{\natexlab{a}})\citenamefont{Xiao,
  Guo, Meng, Jing, and Yung}}]{PhysRevA.100.032118}
\bibinfo{author}{\bibfnamefont{Y.}~\bibnamefont{Xiao}},
  \bibinfo{author}{\bibfnamefont{C.}~\bibnamefont{Guo}},
  \bibinfo{author}{\bibfnamefont{F.}~\bibnamefont{Meng}},
  \bibinfo{author}{\bibfnamefont{N.}~\bibnamefont{Jing}}, \bibnamefont{and}
  \bibinfo{author}{\bibfnamefont{M.-H.} \bibnamefont{Yung}},
  \bibinfo{journal}{Phys. Rev. A} \textbf{\bibinfo{volume}{100}},
  \bibinfo{pages}{032118} (\bibinfo{year}{2019}{\natexlab{a}}),
  \urlprefix\url{https://link.aps.org/doi/10.1103/PhysRevA.100.032118}.

\bibitem[{\citenamefont{Giovannetti et~al.}(2004)\citenamefont{Giovannetti,
  Lloyd, and Maccone}}]{doi:10.1126/science.1104149}
\bibinfo{author}{\bibfnamefont{V.}~\bibnamefont{Giovannetti}},
  \bibinfo{author}{\bibfnamefont{S.}~\bibnamefont{Lloyd}}, \bibnamefont{and}
  \bibinfo{author}{\bibfnamefont{L.}~\bibnamefont{Maccone}},
  \bibinfo{journal}{Science} \textbf{\bibinfo{volume}{306}},
  \bibinfo{pages}{1330} (\bibinfo{year}{2004}),
  \eprint{https://www.science.org/doi/pdf/10.1126/science.1104149},
  \urlprefix\url{https://www.science.org/doi/abs/10.1126/science.1104149}.

\bibitem[{\citenamefont{Giovannetti et~al.}(2006)\citenamefont{Giovannetti,
  Lloyd, and Maccone}}]{PhysRevLett.96.010401}
\bibinfo{author}{\bibfnamefont{V.}~\bibnamefont{Giovannetti}},
  \bibinfo{author}{\bibfnamefont{S.}~\bibnamefont{Lloyd}}, \bibnamefont{and}
  \bibinfo{author}{\bibfnamefont{L.}~\bibnamefont{Maccone}},
  \bibinfo{journal}{Phys. Rev. Lett.} \textbf{\bibinfo{volume}{96}},
  \bibinfo{pages}{010401} (\bibinfo{year}{2006}),
  \urlprefix\url{https://link.aps.org/doi/10.1103/PhysRevLett.96.010401}.

\bibitem[{\citenamefont{Roy and Braunstein}(2008)}]{PhysRevLett.100.220501}
\bibinfo{author}{\bibfnamefont{S.~M.} \bibnamefont{Roy}} \bibnamefont{and}
  \bibinfo{author}{\bibfnamefont{S.~L.} \bibnamefont{Braunstein}},
  \bibinfo{journal}{Phys. Rev. Lett.} \textbf{\bibinfo{volume}{100}},
  \bibinfo{pages}{220501} (\bibinfo{year}{2008}),
  \urlprefix\url{https://link.aps.org/doi/10.1103/PhysRevLett.100.220501}.

\bibitem[{\citenamefont{Giovannetti et~al.}(2011)\citenamefont{Giovannetti,
  Lloyd, and Maccone}}]{Giovannetti2011}
\bibinfo{author}{\bibfnamefont{V.}~\bibnamefont{Giovannetti}},
  \bibinfo{author}{\bibfnamefont{S.}~\bibnamefont{Lloyd}}, \bibnamefont{and}
  \bibinfo{author}{\bibfnamefont{L.}~\bibnamefont{Maccone}},
  \bibinfo{journal}{Nature Photonics} \textbf{\bibinfo{volume}{5}},
  \bibinfo{pages}{222} (\bibinfo{year}{2011}),
  \urlprefix\url{https://doi.org/10.1038/nphoton.2011.35}.

\bibitem[{\citenamefont{Wehner and Winter}(2010)}]{Wehner_2010}
\bibinfo{author}{\bibfnamefont{S.}~\bibnamefont{Wehner}} \bibnamefont{and}
  \bibinfo{author}{\bibfnamefont{A.}~\bibnamefont{Winter}},
  \bibinfo{journal}{New Journal of Physics} \textbf{\bibinfo{volume}{12}},
  \bibinfo{pages}{025009} (\bibinfo{year}{2010}),
  \urlprefix\url{https://doi.org/10.1088/1367-2630/12/2/025009}.

\bibitem[{\citenamefont{Bia{\l}ynicki-Birula and Mycielski}(1975)}]{CMP1975}
\bibinfo{author}{\bibfnamefont{I.}~\bibnamefont{Bia{\l}ynicki-Birula}}
  \bibnamefont{and}
  \bibinfo{author}{\bibfnamefont{J.}~\bibnamefont{Mycielski}},
  \bibinfo{journal}{Communications in Mathematical Physics}
  \textbf{\bibinfo{volume}{44}}, \bibinfo{pages}{129} (\bibinfo{year}{1975}),
  \urlprefix\url{https://doi.org/10.1007/BF01608825}.

\bibitem[{\citenamefont{Deutsch}(1983)}]{PhysRevLett.50.631}
\bibinfo{author}{\bibfnamefont{D.}~\bibnamefont{Deutsch}},
  \bibinfo{journal}{Phys. Rev. Lett.} \textbf{\bibinfo{volume}{50}},
  \bibinfo{pages}{631} (\bibinfo{year}{1983}),
  \urlprefix\url{https://link.aps.org/doi/10.1103/PhysRevLett.50.631}.

\bibitem[{\citenamefont{Kraus}(1987)}]{PhysRevD.35.3070}
\bibinfo{author}{\bibfnamefont{K.}~\bibnamefont{Kraus}},
  \bibinfo{journal}{Phys. Rev. D} \textbf{\bibinfo{volume}{35}},
  \bibinfo{pages}{3070} (\bibinfo{year}{1987}),
  \urlprefix\url{https://link.aps.org/doi/10.1103/PhysRevD.35.3070}.

\bibitem[{\citenamefont{Maassen and Uffink}(1988)}]{PhysRevLett.60.1103}
\bibinfo{author}{\bibfnamefont{H.}~\bibnamefont{Maassen}} \bibnamefont{and}
  \bibinfo{author}{\bibfnamefont{J.~B.~M.} \bibnamefont{Uffink}},
  \bibinfo{journal}{Phys. Rev. Lett.} \textbf{\bibinfo{volume}{60}},
  \bibinfo{pages}{1103} (\bibinfo{year}{1988}),
  \urlprefix\url{https://link.aps.org/doi/10.1103/PhysRevLett.60.1103}.

\bibitem[{\citenamefont{Berta et~al.}(2010)\citenamefont{Berta, Christandl,
  Colbeck, Renes, and Renner}}]{Berta2010}
\bibinfo{author}{\bibfnamefont{M.}~\bibnamefont{Berta}},
  \bibinfo{author}{\bibfnamefont{M.}~\bibnamefont{Christandl}},
  \bibinfo{author}{\bibfnamefont{R.}~\bibnamefont{Colbeck}},
  \bibinfo{author}{\bibfnamefont{J.~M.} \bibnamefont{Renes}}, \bibnamefont{and}
  \bibinfo{author}{\bibfnamefont{R.}~\bibnamefont{Renner}},
  \bibinfo{journal}{Nature Physics} \textbf{\bibinfo{volume}{6}},
  \bibinfo{pages}{659} (\bibinfo{year}{2010}),
  \urlprefix\url{https://doi.org/10.1038/nphys1734}.

\bibitem[{\citenamefont{Coles and Piani}(2014)}]{PhysRevA.89.022112}
\bibinfo{author}{\bibfnamefont{P.~J.} \bibnamefont{Coles}} \bibnamefont{and}
  \bibinfo{author}{\bibfnamefont{M.}~\bibnamefont{Piani}},
  \bibinfo{journal}{Phys. Rev. A} \textbf{\bibinfo{volume}{89}},
  \bibinfo{pages}{022112} (\bibinfo{year}{2014}),
  \urlprefix\url{https://link.aps.org/doi/10.1103/PhysRevA.89.022112}.

\bibitem[{\citenamefont{Xiao et~al.}(2016{\natexlab{b}})\citenamefont{Xiao,
  Jing, Fei, Li, Li-Jost, Ma, and Wang}}]{PhysRevA.93.042125}
\bibinfo{author}{\bibfnamefont{Y.}~\bibnamefont{Xiao}},
  \bibinfo{author}{\bibfnamefont{N.}~\bibnamefont{Jing}},
  \bibinfo{author}{\bibfnamefont{S.-M.} \bibnamefont{Fei}},
  \bibinfo{author}{\bibfnamefont{T.}~\bibnamefont{Li}},
  \bibinfo{author}{\bibfnamefont{X.}~\bibnamefont{Li-Jost}},
  \bibinfo{author}{\bibfnamefont{T.}~\bibnamefont{Ma}}, \bibnamefont{and}
  \bibinfo{author}{\bibfnamefont{Z.-X.} \bibnamefont{Wang}},
  \bibinfo{journal}{Phys. Rev. A} \textbf{\bibinfo{volume}{93}},
  \bibinfo{pages}{042125} (\bibinfo{year}{2016}{\natexlab{b}}),
  \urlprefix\url{https://link.aps.org/doi/10.1103/PhysRevA.93.042125}.

\bibitem[{\citenamefont{Xiao et~al.}(2016{\natexlab{c}})\citenamefont{Xiao,
  Jing, Fei, and Li-Jost}}]{Xiao_2016}
\bibinfo{author}{\bibfnamefont{Y.}~\bibnamefont{Xiao}},
  \bibinfo{author}{\bibfnamefont{N.}~\bibnamefont{Jing}},
  \bibinfo{author}{\bibfnamefont{S.-M.} \bibnamefont{Fei}}, \bibnamefont{and}
  \bibinfo{author}{\bibfnamefont{X.}~\bibnamefont{Li-Jost}},
  \bibinfo{journal}{Journal of Physics A: Mathematical and Theoretical}
  \textbf{\bibinfo{volume}{49}}, \bibinfo{pages}{49LT01}
  (\bibinfo{year}{2016}{\natexlab{c}}),
  \urlprefix\url{https://doi.org/10.1088/1751-8113/49/49/49lt01}.

\bibitem[{\citenamefont{Xiao et~al.}(2017)\citenamefont{Xiao, Jing, and
  Li-Jost}}]{Xiao2016U}
\bibinfo{author}{\bibfnamefont{Y.}~\bibnamefont{Xiao}},
  \bibinfo{author}{\bibfnamefont{N.}~\bibnamefont{Jing}}, \bibnamefont{and}
  \bibinfo{author}{\bibfnamefont{X.}~\bibnamefont{Li-Jost}},
  \bibinfo{journal}{Quantum Information Processing}
  \textbf{\bibinfo{volume}{16}}, \bibinfo{pages}{104} (\bibinfo{year}{2017}),
  \urlprefix\url{https://doi.org/10.1007/s11128-017-1554-6}.

\bibitem[{\citenamefont{Xiao et~al.}(2019{\natexlab{b}})\citenamefont{Xiao,
  Fang, and Gour}}]{xiao2019complementary}
\bibinfo{author}{\bibfnamefont{Y.}~\bibnamefont{Xiao}},
  \bibinfo{author}{\bibfnamefont{K.}~\bibnamefont{Fang}}, \bibnamefont{and}
  \bibinfo{author}{\bibfnamefont{G.}~\bibnamefont{Gour}},
  \emph{\bibinfo{title}{The complementary information principle of quantum
  mechanics}} (\bibinfo{year}{2019}{\natexlab{b}}), \eprint{1908.07694},
  \urlprefix\url{https://arxiv.org/abs/1908.07694}.

\bibitem[{\citenamefont{Xiao et~al.}(2021)\citenamefont{Xiao, Sengupta, Yang,
  and Gour}}]{PhysRevResearch.3.023077}
\bibinfo{author}{\bibfnamefont{Y.}~\bibnamefont{Xiao}},
  \bibinfo{author}{\bibfnamefont{K.}~\bibnamefont{Sengupta}},
  \bibinfo{author}{\bibfnamefont{S.}~\bibnamefont{Yang}}, \bibnamefont{and}
  \bibinfo{author}{\bibfnamefont{G.}~\bibnamefont{Gour}},
  \bibinfo{journal}{Phys. Rev. Research} \textbf{\bibinfo{volume}{3}},
  \bibinfo{pages}{023077} (\bibinfo{year}{2021}),
  \urlprefix\url{https://link.aps.org/doi/10.1103/PhysRevResearch.3.023077}.

\bibitem[{\citenamefont{Huang et~al.}(2018)\citenamefont{Huang, Gan, Xiao, Shu,
  and Yung}}]{Huang2018}
\bibinfo{author}{\bibfnamefont{J.-L.} \bibnamefont{Huang}},
  \bibinfo{author}{\bibfnamefont{W.-C.} \bibnamefont{Gan}},
  \bibinfo{author}{\bibfnamefont{Y.}~\bibnamefont{Xiao}},
  \bibinfo{author}{\bibfnamefont{F.-W.} \bibnamefont{Shu}}, \bibnamefont{and}
  \bibinfo{author}{\bibfnamefont{M.-H.} \bibnamefont{Yung}},
  \bibinfo{journal}{The European Physical Journal C}
  \textbf{\bibinfo{volume}{78}}, \bibinfo{pages}{545} (\bibinfo{year}{2018}),
  \urlprefix\url{https://doi.org/10.1140/epjc/s10052-018-6026-3}.

\bibitem[{\citenamefont{Qian et~al.}(2020)\citenamefont{Qian, Wu, Ji, Xiao, and
  Sanders}}]{PhysRevD.102.096009}
\bibinfo{author}{\bibfnamefont{C.}~\bibnamefont{Qian}},
  \bibinfo{author}{\bibfnamefont{Y.-D.} \bibnamefont{Wu}},
  \bibinfo{author}{\bibfnamefont{J.-W.} \bibnamefont{Ji}},
  \bibinfo{author}{\bibfnamefont{Y.}~\bibnamefont{Xiao}}, \bibnamefont{and}
  \bibinfo{author}{\bibfnamefont{B.~C.} \bibnamefont{Sanders}},
  \bibinfo{journal}{Phys. Rev. D} \textbf{\bibinfo{volume}{102}},
  \bibinfo{pages}{096009} (\bibinfo{year}{2020}),
  \urlprefix\url{https://link.aps.org/doi/10.1103/PhysRevD.102.096009}.

\bibitem[{\citenamefont{Candes et~al.}(2006)\citenamefont{Candes, Romberg, and
  Tao}}]{1580791}
\bibinfo{author}{\bibfnamefont{E.}~\bibnamefont{Candes}},
  \bibinfo{author}{\bibfnamefont{J.}~\bibnamefont{Romberg}}, \bibnamefont{and}
  \bibinfo{author}{\bibfnamefont{T.}~\bibnamefont{Tao}}, \bibinfo{journal}{IEEE
  Transactions on Information Theory} \textbf{\bibinfo{volume}{52}},
  \bibinfo{pages}{489} (\bibinfo{year}{2006}).

\bibitem[{\citenamefont{Hofmann and Takeuchi}(2003)}]{PhysRevA.68.032103}
\bibinfo{author}{\bibfnamefont{H.~F.} \bibnamefont{Hofmann}} \bibnamefont{and}
  \bibinfo{author}{\bibfnamefont{S.}~\bibnamefont{Takeuchi}},
  \bibinfo{journal}{Phys. Rev. A} \textbf{\bibinfo{volume}{68}},
  \bibinfo{pages}{032103} (\bibinfo{year}{2003}),
  \urlprefix\url{https://link.aps.org/doi/10.1103/PhysRevA.68.032103}.

\bibitem[{\citenamefont{Rutkowski et~al.}(2017)\citenamefont{Rutkowski,
  Buraczewski, Horodecki, and Stobi\ifmmode~\acute{n}\else
  \'{n}\fi{}ska}}]{PhysRevLett.118.020402}
\bibinfo{author}{\bibfnamefont{A.}~\bibnamefont{Rutkowski}},
  \bibinfo{author}{\bibfnamefont{A.}~\bibnamefont{Buraczewski}},
  \bibinfo{author}{\bibfnamefont{P.}~\bibnamefont{Horodecki}},
  \bibnamefont{and}
  \bibinfo{author}{\bibfnamefont{M.}~\bibnamefont{Stobi\ifmmode~\acute{n}\else
  \'{n}\fi{}ska}}, \bibinfo{journal}{Phys. Rev. Lett.}
  \textbf{\bibinfo{volume}{118}}, \bibinfo{pages}{020402}
  (\bibinfo{year}{2017}),
  \urlprefix\url{https://link.aps.org/doi/10.1103/PhysRevLett.118.020402}.

\bibitem[{\citenamefont{Xiao et~al.}(2020)\citenamefont{Xiao, Xiang, He, and
  Sanders}}]{Xiao_2020}
\bibinfo{author}{\bibfnamefont{Y.}~\bibnamefont{Xiao}},
  \bibinfo{author}{\bibfnamefont{Y.}~\bibnamefont{Xiang}},
  \bibinfo{author}{\bibfnamefont{Q.}~\bibnamefont{He}}, \bibnamefont{and}
  \bibinfo{author}{\bibfnamefont{B.~C.} \bibnamefont{Sanders}},
  \bibinfo{journal}{New Journal of Physics} \textbf{\bibinfo{volume}{22}},
  \bibinfo{pages}{073063} (\bibinfo{year}{2020}),
  \urlprefix\url{https://doi.org/10.1088/1367-2630/ab9d57}.

\bibitem[{\citenamefont{Coles et~al.}(2017)\citenamefont{Coles, Berta,
  Tomamichel, and Wehner}}]{RevModPhys.89.015002}
\bibinfo{author}{\bibfnamefont{P.~J.} \bibnamefont{Coles}},
  \bibinfo{author}{\bibfnamefont{M.}~\bibnamefont{Berta}},
  \bibinfo{author}{\bibfnamefont{M.}~\bibnamefont{Tomamichel}},
  \bibnamefont{and} \bibinfo{author}{\bibfnamefont{S.}~\bibnamefont{Wehner}},
  \bibinfo{journal}{Rev. Mod. Phys.} \textbf{\bibinfo{volume}{89}},
  \bibinfo{pages}{015002} (\bibinfo{year}{2017}),
  \urlprefix\url{https://link.aps.org/doi/10.1103/RevModPhys.89.015002}.

\bibitem[{\citenamefont{Xiao et~al.}(2022)\citenamefont{Xiao, Yang, Wang, Liu,
  and Gu}}]{Xiao2022}
\bibinfo{author}{\bibfnamefont{Y.}~\bibnamefont{Xiao}},
  \bibinfo{author}{\bibfnamefont{Y.}~\bibnamefont{Yang}},
  \bibinfo{author}{\bibfnamefont{X.}~\bibnamefont{Wang}},
  \bibinfo{author}{\bibfnamefont{Q.}~\bibnamefont{Liu}}, \bibnamefont{and}
  \bibinfo{author}{\bibfnamefont{M.}~\bibnamefont{Gu}}, \bibinfo{journal}{Under
  preparation}  (\bibinfo{year}{2022}).

\bibitem[{\citenamefont{Mondal et~al.}(2017)\citenamefont{Mondal, Bagchi, and
  Pati}}]{PhysRevA.95.052117}
\bibinfo{author}{\bibfnamefont{D.}~\bibnamefont{Mondal}},
  \bibinfo{author}{\bibfnamefont{S.}~\bibnamefont{Bagchi}}, \bibnamefont{and}
  \bibinfo{author}{\bibfnamefont{A.~K.} \bibnamefont{Pati}},
  \bibinfo{journal}{Phys. Rev. A} \textbf{\bibinfo{volume}{95}},
  \bibinfo{pages}{052117} (\bibinfo{year}{2017}),
  \urlprefix\url{https://link.aps.org/doi/10.1103/PhysRevA.95.052117}.

\bibitem[{\citenamefont{Huang}(2012)}]{PhysRevA.86.024101}
\bibinfo{author}{\bibfnamefont{Y.}~\bibnamefont{Huang}},
  \bibinfo{journal}{Phys. Rev. A} \textbf{\bibinfo{volume}{86}},
  \bibinfo{pages}{024101} (\bibinfo{year}{2012}),
  \urlprefix\url{https://link.aps.org/doi/10.1103/PhysRevA.86.024101}.

\bibitem[{\citenamefont{Yu et~al.}(2019)\citenamefont{Yu, Jing, and
  Li-Jost}}]{PhysRevA.100.022116}
\bibinfo{author}{\bibfnamefont{B.}~\bibnamefont{Yu}},
  \bibinfo{author}{\bibfnamefont{N.}~\bibnamefont{Jing}}, \bibnamefont{and}
  \bibinfo{author}{\bibfnamefont{X.}~\bibnamefont{Li-Jost}},
  \bibinfo{journal}{Phys. Rev. A} \textbf{\bibinfo{volume}{100}},
  \bibinfo{pages}{022116} (\bibinfo{year}{2019}),
  \urlprefix\url{https://link.aps.org/doi/10.1103/PhysRevA.100.022116}.

\end{thebibliography}

\end{document}